\documentclass[11pt,twoside]{article}

\usepackage{asp2014}

\aspSuppressVolSlug
\resetcounters

\begin{document}

\title{Agile Software Engineering and Systems Engineering at SKA Scale}

\author{Juande Santander-Vela$^1$
\affil{$^1$SKA Organisation, Lower Withington, Cheshire, UK; \email{j.santander-vela@skatelescope.org}}
}

\paperauthor{Juande Santander-Vela}{j.santander-vela@skatelescope.org}{orcid.org/0000-0002-2660-510X}{SKA Organisation}{Project Engineering}{Lower Withington}{Cheshire}{SK11 9DL}{United Kingdom}

\begin{abstract}
Systems Engineering (SE) is the set of processes and documentation required for successfully realising large-scale engineering projects, but the classical approach is not a good fit for software-intensive projects, especially when the needs of the different stakeholders are not fully known from the beginning, and requirement priorities might change. The SKA is the ultimate software-enabled telescope, with enormous amounts of computing hardware and software required to perform its data reduction. We give an overview of the system and software engineering processes in the SKA1 development, and the tension between classical and agile SE.
\end{abstract}


\section{Introduction} 
\label{sec:introduction}
The Square Kilometre Array (SKA) is an international project with the aim of building multi-purpose radio telescopes, with an equivalent collecting area of at least one square kilometre. It will provide unprecedented sensitivity and survey speed, so that key questions in modern astrophysics and cosmology can be answered. Currently, Phase 1 of the SKA (SKA1) is under design, nearing Critical Design Review (CDR) status.	

The data rates generated by each of the SKA1 telescopes is in the order of 1 terabyte/s at the interface with the Central Signal Processor (CSP). The amount of processing required by the CSP to deal with that traffic is ~50\% of the current most powerful supercomputer in the world,\footnote{98 PFLOPS as of June 2017, see \url{https://top500.org}} even if they are simple Multiply-Accumulate operations, and the Science Data Processor (SDP) will require in the order of 2.5 times the capability of the current most powerful computer in the world. 

Eleven international Consortia, whose members span 17 timezones,\footnote{From Vancouver (GMT-9) to Sydney (GMT+12); if we include New Zealand, and the difference between daylight savings in northern and southern hemispheres, we can reach 19-21 hours; the sun always shines over at least one SKA contributor.}A are designing each of the subsystems.
However, the SKA Organisation (SKAO) is the design authority, with Integrated Engineering Teams (IETs) per Element with Systems Engineering, Project Management, Operations, and Science considerations always represented. The External IETs include the counterparts from Consortia, and we have Telescope Teams which can spawn Resolution Teams when issues need to be discussed and resolved across the whole project.
SKAO retains ownership of the L1 requirements, System Architecture, System Baseline Design, and Functional Analysis, and after CDR we will have ownership of all system and subsystem requirements, and the SE and design artefacts from Consortia.
More information of SKA System Engineering practices and challenges can be found in~\citep{2016SPIE.9911E..0WC}.


\section{Systems Engineering vs Agile development principles} 
\label{sec:se_vs_agile_development_principles}
Systems Engineering (SE) is an interdisciplinary approach and means to enable the full life cycle of successful systems, including problem formulation, solution development, and operational sustainment and use~\citep{INCOSE-TP-2003-002-04}. During all of those phases, understanding of the system increases, and changes are inevitable. \emph{Managing change}, and making sure change happens while keeping a working system, is central to SE.

However, in most of the typical SE processes there is the underlying assumption that \emph{change will decrease with time}, which is typically true for hardware/mechanical systems, but \emph{tends not to be true for software heavy systems}. 

The main principles from the Agile Manifesto\footnote{\url{http://agilemanifesto.org}} are:

\begin{itemize}
	\item \emph{Individuals and interactions} over processes and tools;
	\item \emph{Working software} over comprehensive documentation;
	\item \emph{Customer collaboration} over contract negotiation;
	\item \emph{Responding to change} over following a plan
\end{itemize}

Following those principles, the traditional Software Engineering Management Plan becomes more a framework for the best practices that need to be followed ---pull requests between branches/forks versus commits to master, code reviews, Test Driven Design---, and not a final blueprint on how the software will be managed.

It also means that traditional Work-Breakdown-Structure oriented approaches don't fully work, and that instead, \emph{what is finally built \textbf{at each stage} is a result of agreements with stakeholders}, as we will:
\begin{itemize}
	\item Keep \emph{communication with all stakeholders} involved,with speedy and clear dissemination of agreements;
	\item To have \emph{continuous integration of engineering artifacts}, and software;
	\item To \emph{hire software developers (or development entities) that are invested in the system}, not merely contracted; but also sharing gains with contractors if something is done faster than expected, or some desirable metrics are exceeded	
	\item To \emph{have systems and processes that accommodate system (and software) change} and can still \emph{prove requirements are satisfied}; this means \emph{not using a document-centric requirements management system}, but being able to \emph{integrate requirements, development, testing, and results} with a quick turnaround.	
\end{itemize}

And all of this is made on a \emph{fixed cadence} ---with the fastest period being the \emph{sprint}---, trying to keep momentum, allowing teams to self-organise their prioritised tasks.

We must not forget that the reason to become agile is to embed quality in the system being developed, and that the system is useful to the different stakeholders, which is also a goal of SE. \citet{I10-1_adassxxvii} expands on quality and software.


\section{Marrying SE with Agile} 
\label{sec:se_and_agile}
Successfully marrying Agile principles and SE principles means that, instead of making sure the SE processes are correct so that we remove the human factor as a source of mistakes, and that we can prove that the system, as designed, will behave as intended when built, \emph{we focus on delivering a minimum viable system as soon as possible, constructed from minimum viable subsystems, and we keep adding functionality}, making sure we keep the intended behaviours that we have already built, and that we incorporate the still not incorporated behaviours as development progresses.	

That means that we need metrics to keep track of the progress towards design completion during the design phase ---the downward part of the classical V---, and then to record validation of the design artefacts ---upward part of the V---. But you also want to be able to support those processes even when the design phase does not go through a traditional V, but it is made of multiple, indented Vs, or can be evolved, and the processes followed for each potential delivery.

If the killer function of Agile development is Continuous Integration, and be always available to deliver working software, \emph{the killer function for Agile SE is Continuous Inspection}, and be always able to deliver status of compliance, and of progress, almost at any moment, and on cadence.	


\section{SE functions during SKA pre-construction and construction} 
\label{sec:se_functions_during_ska_pre_construction}
During pre-construction, the main SE function from SKAO is the supervision of Interface Control Documents (ICDs), but also assessing design compliance against L2 requirements, through compliance matrices, and evidence for compliance. Many statements can be easily justified through functional allocations, while others require architectural support in order for them to be met.

Given the different levels of design maturity at system and elevent level, and of adherence to SE principles across Consortia, and the relatively small amounts of SE personell at SKAO, tradicional processes with full flown-down from Design Reference Mission, to Operational Concept, to Functional Analysis, to System Requierements have not been fully possible. Rather, iterations over the available detail of those SE artefacts have been steadily increasing definition, with the drawing of Element and System PDR baselines, and future Element and System CDR baselines.

In all these processes, we have required submissions of some design and SE artifacts linked with individual milestones. SKAO has a unified system for Configuration Management, eBentley\footnote{\url{https://www.bentley.com/en/products/product-line/asset-performance/assetwise-alim}} (eB), that is also used by all Consortia. However, in spite of SKAO offers to Consortia for usage of our own Requirements Management system (Jama\footnote{\url{http://jamasoftware.com/}}), it has not been standardised, and requirement submissions require manual processes from the SKAO SEs. There is still a large amount of engineering understanding required, and the process enables a less mechanistic, more engineering-centric review of requirements, but it is also more prone to pitfalls.

For SKA1 construction, the Scaled Agile Framework\footnote{\url{http://scaledagileframework.com}} (SAFe) has been selected as the way software development will be performed. SAFe allows for recursive groupings of Agile teams (using Scrum) into Agile Release Trains (ARTs) ---that also use Scrum for managing the Train backlog, that then gets assigned to the ART backlog(s)---, and can group multiple ART into a Large Solution.
As previously indicated, the SE function will be organised from the SKA Inter-Governmental Organisation (IGO) that will exist before triggering construction. However, it needs to trickle down to the Agile Release Trains, and to the individual agile teams.	
Part of it has to go through the figure of the product owners, and consumer representatives, but there must be a strong presence of the system's mission and behaviour, and the Operations Concept Document, so that it is really implemented. We inted to define additional metrics from the inspection of the backlogs, and the processing of the outcomes of the Program Increment demos.	


\section{Conclusions} 
\label{sec:conclusions}
Applying Systems Engineering \emph{by the book} is never a good idea. It is even less ideal for projects that are optimising a multi-valued function in the cost, science capability, and time-to-construction, among other dimensions. Processes need to be tailored, and they need to be kept responsive, while at the same time keeping track of the SE deliverables that are required as design artefacts, and managing the risk of the uncertainties that are still carried forward to the construction phase.
When software is involved, design needs to focus on documenting the system architecture, instead of designing down to the application or module level, as agile methods will require continuous adaptation of the modules to be developed. 
It also requires definition and constant evaluation of the key performance metrics that will be used during the construction phase, and be able to translate between SE artefacts such as ICDs, the software architecture, functional analysis, and the needs for the different releases.	
We have been applying evidence-based SE artefacts by different means, and found the shortcomings of the approach, specially when dealing with a world-wide distributed team with no traditional line of command.



\bibliography{O4-4}  

\begin{thebibliography}{}
\expandafter\ifx\csname natexlab\endcsname\relax\def\natexlab#1{#1}\fi
\expandafter\ifx\csname url\endcsname\relax
  \def\url#1{\texttt{#1}}\fi
\expandafter\ifx\csname urlprefix\endcsname\relax\def\urlprefix{URL }\fi
\providecommand{\eprint}[2][]{\url{#2}}

\bibitem[{{Cremonini} et~al.(2016){Cremonini}, {Caiazzo}, {Hayden}, {Labate},
  {Olgin}, \& {Santander-Vela}}]{2016SPIE.9911E..0WC}
{Cremonini}, A., {Caiazzo}, M., {Hayden}, D., {Labate}, M.~G., {Olgin}, R., \&
  {Santander-Vela}, J. 2016, in Modeling, Systems Engineering, and Project
  Management for Astronomy VI, vol. 9911 of Proc. SPIE, 99110W

\bibitem[{INCOSE(2015)}]{INCOSE-TP-2003-002-04}
INCOSE 2015, Systems Engineering Handbook (Wiley)

\bibitem[{Rees(2018)}]{I10-1_adassxxvii}
Rees, N. 2018, in ADASS XXVII, edited by TBD (San Francisco: ASP), vol. TBD of
  ASP Conf. Ser., TBD

\end{thebibliography}

\end{document}